\documentstyle[twocolumn,aps]{revtex}
\begin{document}
\draft
\title{Effect of Decoherence on Bell's Inequality for an EPR Pair
\thanks{SPS-JNU preprint}}

\author{A. Venugopalan, Deepak Kumar and R. Ghosh}

\address{School of Physical Sciences,
Jawaharlal Nehru University,
New Delhi-110 067, India}
\maketitle
\begin{abstract}

According to Bell's theorem, the degree of correlation between
spatially separated measurements on a quantum system is limited
by certain inequalities if one assumes the condition of locality.
Quantum mechanics predicts that this limit can be exceeded,
making it nonlocal. We analyse the effect of an environment
modelled by a fluctuating magnetic field on the quantum correlations
in an EPR singlet as seen in the Bell inequality. We show that
in an EPR setup, the system goes from the usual 'violation'
of Bell inequality to a 'non-violation' for times larger than
a characteristic time scale which is related to the parameters
of the fluctuating field. We also look at these inequalities
as a function of the spatial separation between the EPR pair.
\end{abstract}


\section{Introduction}

The basic formalism of quantum theory which involves concepts
of probability amplitudes and the superposition principle has
been the source of many paradoxes and strange implications.
A celebrated paradox that illustrates the counterintuitive conceptual
framework of quantum mechanics is the 'E-P-R paradox' which
is based on a gedankenexperiment proposed by Einstein, Podolsky
and Rosen $ ( $ E-P-R) in 1935 [1].

E-P-R showed that quantum theory is incomplete as the quantum
description of the physical system does not contain all the
'elements of reality'. The so-called 'elements of reality' are
those physical attributes of the system that can be measured
without disturbing the system. E-P-R's argument also required
that the principle of locality be obeyed in the sense that the
disturbance caused by the act of measurement does not travel
faster than light. It seems intuitive to expect all physical
theories to be local and consistent with realism as stated above.
But E-P-R's argument showed that quantum mechanics is inconsistent
with the doctrines of realism and locality. A natural requirement
at that time was, thus, to look for a theory that agreed with
all the predictions of quantum mechanics and yet did not share
its conflicts with realism and locality. One possibility was
to reinterpret quantum mechanics in terms of a statistical account
of an underlying hidden-variable theory which would conform
to locality and realism. Many attempts were made to find such
a complete 'hidden variable theory'. In 1965 Bell proved that
nonlocal correlations among measured quantities must obey
some general inequalities that are common to all local realistic
theories [2]. Quantum mechanics, on the other hand, gives rise
to correlations that violate Bell's inequalities. This provided
definite experimental ways to test for any hidden-variable theories
that may underlie quantum mechanics. A number of experiments
have been carried out to measure such correlations and most
of them have yielded results which are in excellent agreement
with quantum theory and in disagreement with local realistic
theories [3,4]

In the following we describe Bohm's version [5,6] of the
E-P-R gedankenexperiment which was considered by Bell to prove
his theorem. A bound state of two spin-1/2 particles is prepared
in the quantum-mechanical singlet state $ \mid \psi \rangle $
given by

\begin{equation}
 \mid \psi \rangle = { 1\over \sqrt{ 2}}\biggl(\mid \uparrow_{1n
}\rangle
\otimes \mid \downarrow_{2n }\rangle - \mid \downarrow_{1n }\rangle \otimes
\mid \uparrow_{2n }\rangle
\biggr) ,
\end{equation}
where $ \mid \uparrow_{in }\rangle $ and $ \mid \downarrow_{in }\rangle , i
= 1 , 2 ,
$ describe the spin states in which particle $ i $ has spin 'up'
and 'down', respectively, along the direction $ n . $ The
bound state is now broken in such a way that the particles fly
apart in opposite directions, but their total spin state remains
a singlet. Two detectors, far apart from each other $ , $
are arranged to measure the spin component of particle $ 1 $
along a direction a and that of particle $ 2 $
along a direction $ b . $ The directions a and
$ b $ are at an angle $ \theta $ with respect to each
other. According to the quantum formalism, the measured expectation
value can be written as

\begin{equation}
 E( a , b ) \equiv \langle\psi \mid \sigma_{1 }. a\sigma_{2 }. b\mid \psi
\rangle = -cos\theta .
\end{equation}
One can see that for $ \theta = $ 0, E( a $ , b )
= $ -1. This implies the existence of a perfect negative correlation
between the spins of $ 1 $ and $ 2 . $ Thus, one
can predict with certainty the result of a measurement on $ 2
 $ by previously obtaining the result of $ 1 . $ This
measurement on $ 1 $ does not disturb the state of the
particle $ 2 , $ which is very far from $ 1 .
$ Therefore, one argues that all components of the spin of $ 2
 $ are 'elements of reality', which quantum mechanics cannot
describe. Further, since the quantum mechanical state $ \mid \psi \rangle $
does not determine the result of an individual measurement,
there must exist a more complete specification of the state
through a hidden-variable theory. By assuming a suitable condition
of locality, Bell [2] showed that for any local, realistic,
deterministic theory, the expectation value (2) must satisfy
a simple inequality, one form of which is

\begin{equation}
 \mid E( a , b )-E( a , c )\mid \le 1 + E( b , c ),
\end{equation}
where a $ , b $ and $ c $ are three directions
along which the detectors measure the spin components of $ 1 $
and $ 2 . $ One can easily see that for the condition $ \theta ( $
a $ , b ) =\theta ( b , c )
= \pi $ /3 and $ \theta ( $ a $ , c ) = 2\pi $ /3,
quantum mechanics clearly violates this inequality, a result
verified by many successful experiments [4, 6-7]. These results
establish nonlocality as an inescapable 'fact of life' and cast
doubts on the notion of 'elements of reality'. There are many
exciting implications of quantum nonlocality like quantum cryptography
and quantum computers. Particularly fascinating is the idea
of 'teleportation' of a quantum state using E-P-R pairs [8].

However, we know from everyday experience that nonlocal correlation
of the E-P-R kind are not seen in the macroscopic physical world.
Classical systems are known to conform to realism and locality.
How does this transition take place? How do nonlocal correlations
disappear and how does classicality emerge? The transition from
quantum to classical is marked by the loss of quantum coherence
which reduces a pure state density matrix for the system to
a classically interpretable diagonal mixture. In recent
years this 'decoherence' [9,10] approach has provided a way
to understand the emergence of classicality from underlying
quantum dynamics. The approach recognizes that a macroscopic
system is never isolated from its environment but is constantly
interacting with it. In this approach one calculates the combined
density matrix of the system and the environment but since one
is interested only in monitoring the system's degrees of freedom,
one traces over the environmental degrees of freedom. Then one
finds that the reduced density matrix of the system is driven
to a diagonal form in time.

In this paper we study the effect of an environment on an E-P-R
singlet where the particles are acted upon by forces which push
them apart in opposite directions. The effect of the environment
here is two-fold. Firstly, the two particles are coupled to
a bath of harmonic oscillators via a Caldeira-Leggett type of
dissipative coupling through their positional degrees of freedom
[9]. This aspect of the model has been discussed in detail elsewhere
[10]. Secondly, the spin degrees of freedom for the two particles
are coupled to an environment modelled by a fluctuating external
magnetic field. We calculate the spin correlations for this
system, and we show that the system goes from a 'violation'
of the inequality at t=0 to a 'nonviolation' over a time scale
related to the parameters of the fluctuating magnetic field.
We also look at this transition as a function of the spatial
separation between the E-P-R pair. In Section II we introduce
the model and set up the equations that the reduced density
matrix of the system obeys. In Section III we obtain the solutions
for these equations and calculate the correlations involved
in the Bell inequality. Finally, in Section IV we summarize
the results of this paper.

\section{The model}

Our model consists of two particles whose initial state
can be described by the following wave function

\begin{equation}
 \psi = \phi ( 1 ) \phi ( 2 )
\otimes { 1\over \sqrt{ 2}} \biggl( \mid \uparrow_{1n }\rangle
\otimes \mid \downarrow_{2n }\rangle - \mid \downarrow_{1
n }\rangle \otimes \mid \uparrow_{2n }\rangle
\biggr),
\end{equation}
where $ \phi ( 1 $ ), $ \phi ( 2 ) $ represent
the spatial parts of the wavefunction which are initially Gaussian
wave packets with zero initial momenta and centered around x=0

\begin{equation}
 \phi (x,0) = { 1\over (d r\pi )^{1/2}} exp \biggl( - x
^{2 }/2d^{2 }\biggr) .
\end{equation}
The particles are coupled to an environment consisting of a
collection of harmonic oscillators via their position coordinates.
Such a model of the environment has been studied in great detail
by many authors in the context of quantum dissipative systems
and the quantum measurement problem [9-10]. In addition to this
coupling, the spin degrees of freedom for the two particles
are coupled to an environment modelled by a fluctuating magnetic
field. The Hamiltonian for this model is:

\begin{eqnarray}
 H &=& { P_{1}^{2}\over 2 m} + { P_{2}^{2}\over 2 m} + \epsilon
X_{1 }- \epsilon X_{2 } \nonumber\\
&&+ \sum { p_{1k}^{2}\over 2 m_{1k}} + { m_{1
k}\omega_{1k}^{2}\over 2} \biggl( x_{1k }- { c_{1
k }X_{1}\over m_{1k}\omega_{1k}^{2}} \biggr)^{2 } \nonumber\\
&&+ \sum{ p_{2k}^{2}\over 2 m_{2k}} + { m_{2k}\omega
_{2k}^{2}\over 2} \biggl( x_{2k }- { c_{2k }X_{2}\over
m_{2k}\omega_{2k}^{2}} \biggr)^{2 }\nonumber\\
&&- { 1\over 2} g\hbar( B_{1 }.\sigma_{1 }+ B_{2 }.\sigma_{2 }),
\end{eqnarray}
where $ P_{1 }, X_{1 }$ and $ P_{2 }, X_{2 }$ are the momentum and position
coordinates of particles $ 1 $ and $ 2$, respectively,
$ \epsilon $ is the strength of the applied force (the force
being positive for $ 1 $ and negative for $ 2 $ ),
$ p_{1k }, x_{1k }$ and $ p_{2k }, x_{2k }$
are the momentum and position coordinates of the $ k^{th
 }$ harmonic oscillator of the baths which couple to $ 1 $ and
$ 2 , c_{1k }$ and $ c_{2k}$ are the respective coupling strengths and
$\omega_{1k }$ and $ \omega_{2k }$ are the frequencies of the oscillators.
$ B_{1 }$ and $B_{2 }$ are the external stochastic magnetic
fields acting on particles $ 1 $ and $ 2 , $ and
$ g $ is the gyromagnetic ratio. Though in our model, we assume
that the external fluctuating fields $ B_{1 }( t )$ and $ B_{2 }( t ) $
are random in time, we envisage the physical origin of this randomness
to arise due to spatial variation of the field in the region where the
singlet is separating. Since the analytical treatment for the
spatially varying field is very difficult, we model the spatial
variation seen by the wavepackets of the particles to be a temporal
variation. $ B_{1 }( t $ ), $ B_{2 }( t )
$ are taken to be Gaussian random processes, having the following
correlations:

\begin{equation}
 g^{2 }\langle B_{i\alpha }( t
_{1 }) B_{i\beta }( t
_{2 })\rangle = { 1\over 2}_{\tau \alpha \alpha }\delta_{\alpha
\beta }\delta ( t
_{1 }- t_{2 }),
\end{equation}
where $ i = $ 1,2 and $ \beta $ refer to the components of the
vector $ B_{i }, \tau_{\alpha \alpha }$ are
the correlation times.. The reduced density matrix of the system,
$ \rho (x_{1 }, s_{1 }, y_{1 },
s_{1}^{ }; x_{2 }, s_{2 },
y_{2 }, s_{2}^{ }) =  \langle x_{1},s_{1}; x_{2 },s_{2}\mid \rho_{R }\mid
y_{1 },s_{1}^{};y_{2 },s_{2}^{ }\rangle $ can
be obtained by using the Feynman-Vernon functional integral
method in which one first writes the path-integral expression
for the complete density matrix and then traces over the oscillator
degrees of freedom. Under conditions of weak coupling to the
oscillator variables, (i.e., the Markovian limit) one finds
that $ \rho_{R }$ obeys an equation which can be
written as

\begin{equation}
 { \partial \rho_{R}\over \partial t} = -i [ L_{s1 }+ L_{s
2 }]\rho_{R }+ i g/2 [ B
_{1 }.\sigma + B_{2 }.\sigma ,
\rho_{R }].
\end{equation}
Here $ L_{s_{1} }$ and $ L_{s_{2} }$ are a form of quantum-Liouville
operators which in spatial representation are given by

\begin{eqnarray}
 L_{sj }&=& { -\hbar \over 2 m}\bigl\{ { \partial^{2}\over
\partial x^{2}_{j}} -{ \partial^{2}\over \partial y^{2}_{j}}
\bigr\} - i\gamma ( x_{j }- y
_{j })\bigl\{{ \partial \over\partial x_{j}} -{ \partial \over\partial
y_{j}}\bigr\} \nonumber\\
&& -{ i D\over 4\hbar^{2}} ( x_{j }- y_{j })^{2}\pm
{i\epsilon \over\hbar }
\end{eqnarray}
where the upper sign is for particle $ 1 $ and the lower
one for particle $ 2 , $ and $ \gamma $ and $ D
 $ are related to the environment, playing the role analogous
to friction and diffusion coefficients, respectively. Since
the particles are noninteracting (apart from the correlation
due to the initial condition) the density matrix equation is
separable, i.e., $ \rho_{R }= \rho_{1
}\otimes \rho_{2 }$ with each of $ \rho_{1
 }$ and $ \rho_{2 }$ satisfying the equation

\begin{equation}
 { \partial \rho_{j}\over \partial t} = -i L_{sj }\rho_{j
}+ { i g\over 2} [ B_{j }.\sigma_{j }, \rho_{j }], j = 1,2.
\end{equation}
We can further effect the separation between the space and the
spin degrees of freedom for each particle by recognizing that
$ \rho_{j }$ is a 2x2 matrix in spin indices and
can therefore be written as
\begin{equation}
 \rho_{j }= { 1\over 2} \biggl(\rho_{sj }( x
_{j }, y_{j }, t ) {\bf 1}
 + W_{j }( x_{j }, y
_{j }, t ).\sigma_{j}\biggr) ,
\end{equation}
where $ {\bf 1} $ denotes a 2x2 unit matrix. Substituting Eq.
(11) in Eq. (10) one easily finds the dynamical equations governing
$ \rho_{sj }$ and $ W_{j }:$
\begin{equation}
 { \partial \rho_{sj}\over \partial t} = -iL_{sj }\rho_{s
j }( x
_{j }, y_{j }, t ),
\end{equation}

\begin{equation}
 { \partial W_{j }( x_{j }, y
_{j }, t)\over  \partial t} = -iL_{sj }W_{j }-g B_{j }x W_{j }.
\end{equation}
{}From Eq.(11) it is clear that $ W_{j }(x_{j }y
_{j }; t ) = \rho_{sj }$ (x
$ _{j}, y_{j }, t ) P_{j
}$ and the polarization vector $ P $ obeys the equation
\begin{equation}
 { \partial P_{j }( t)\over \partial t} = g P_{j }x B
_{j }.
\end{equation}
Thus the complete solution for the problem can be written as
\begin{eqnarray}
 \rho_{R }( x_{1 },s_{1 }, y
_{1 },s_{1}^{ }; x_{2 },s
_{2 }, y_{2 },s_{2}^{ }, t ) &=& { 1\over 4} \rho_{s_{1} }( x_{1
}, y_{1 }, t )\nonumber\\
&& \rho_{s_{2} }( x_{2 }, y_{2 }, t ) \nonumber\\
&&\times [ {\bf 1} + \sigma_{1}. P_{1 }( t )]
 _{s_{1}s_{1}^{\prime} }\nonumber\\
&&[ {\bf 1} +\sigma_{2}. P_{2 }( t )]_{s_{2}s_{2}^{\prime}}
\end{eqnarray}
In the next Section we write down the solutions for equations
(12) and (14) and calculate the correlations involved in the
Bell inequality using Eq. (15).

\section{The density-matrix of the E-P-R pair}

The solution for the spatial part of the density matrix, i.e.,
the solution of Eq. (12) have been worked out in detail elsewhere
[10] $ . $ For the initial condition of Eq. (5), these
solutions are

\begin{eqnarray}
 \rho_{sj }( R_{j }, r
_{j }, t ) &=& \sqrt{ { \pi \over M (\tau )}} exp \bigl\{ - \biggl( {
1\over 4d_{j}^{2}} e^{-2\tau }\nonumber\\
&&+ { D\over 8\hbar^{2}}(1-e^{-2\tau })
 \biggr) r_{j}^{2 }\mp
{ i\epsilon \over\hbar \gamma }( \epsilon^{ \tau }) r_{j}\nonumber\\
&& - { 1\over M (\tau )} \biggl( R_{j }\pm {\epsilon \over m\gamma
^{2}} (1-e^{-\tau }-\tau )\nonumber\\
&& - { i\hbar r_{j}\over 2d_{j}^{2
}m\gamma }\epsilon^{ \tau }(1-e^{-\tau })\nonumber\\
&& - { i D r_{j}\over 4 m\gamma^{2}}_{\hbar }(1-e^{-\tau
})^{2 }\biggr)^{2 }\bigr\}, j = 1, 2
\end{eqnarray}
where the upper signs are for $ j $ =1 and the lower signs are for
j=2, $ \tau = \gamma t , R_{j }=
(x_{j }+ y_{j }$ )/2, $ r_{j
 }= x_{j }- y_{j }, $ and

\begin{eqnarray}
 M (\tau ) &=& d_{j}^{2
 }+ { \hbar^{2}\over d_{j}^{2 }m^{2}\gamma }_{2 }(1 - e^{-\tau
 })^{2}\nonumber\\
&&+ { D\over 2 m^{2}\gamma^{3}} ( 2\tau - 3+4e^{-\tau }-e^{-2\tau }).
\end{eqnarray}
We first note that the off-diagonal part of $ \rho_{s
 j }$ corresponding to $ r_{j }\neq 0
$ vanishes rapidly in time. If we look at the diagonal parts of
(16), i.e., the position distribution functions in the $ t
 \rightarrow \infty $ limit, these solutions represent two Gaussian
wave packets centered around $ + { \epsilon \tau \over m\gamma^{2}} $
and
$ - { \epsilon \tau \over m\gamma^{2}} $ which are moving away from each
other with time.

We now consider Eq. (14). Note that it is exactly the equation
for a classical moment in an external magnetic field. Since
the field $ B ( t ) $ is stochastic, $ P ( t )
$ will also be a stochastic variable. The quantity of interest
corresponding to a real physical observation would, therefore,
be the expectation value of $ P ( t ) ,
$ averaged over the ensemble of the random process $ B ( t ) .
$ One has to, thus, find the solution to a stochastic
Liouville equation $ , $ i.e., an equation for the probability
distribution $ f ( P , t ) $ for the stochastic
variable $ P ( t ) $ [11] $ . $ This problem
has been studied extensively in the literature [11,12] $ .
 $ For the sake of completeness, we record the main results.
The equation obeyed by the probability distribution function
$ f $ (P, $ t ) $ averaged over the ensemble of the
random process is

\begin{equation}
 { \partial f\over \partial t} = - G f ,
\end{equation}
where
\begin{equation}
 G = 1/2 [ L_{x}^{2 }/2\tau_{x
x }+ L_{y}^{2 }/2\tau_{yy
}+ L_{z}^{2 }/2\tau_{zz }],
\end{equation}
and the operator $ L $ is given by

\begin{equation}
 L = -i P x { \partial \over\partial P}
\end{equation}
{}From Eq. (18) one can write down the equation for the average
value of $ P $ as

\begin{equation}
 { \partial \langle P \rangle\over \partial t} = -\langle G^{+ }P \rangle ,
\end{equation}
which reduces to the well-known Bloch equations if we choose
$ \tau_{xx }= \tau_{yy }= \tau_{1 }$ and $ \tau_{zz }= \tau
_{0
}$ where the z-axix is chosen to be the axis of motion of
the E-P-R pair. Eq (21) then yields

\begin{eqnarray}
 { d\langle P_{z }\rangle\over d t} &=& - { 1\over 2\tau_{1 }}\langle P_{z
}\rangle,\eqnum{22a}\\
 { d\langle P_{\pm }\rangle\over d t} &=& -\biggl({ 1\over 4\tau_{1 }}
+ { 1\over 4\tau_{0 }} \biggr) \langle P_{\pm }\rangle . \eqnum{22b}
\end{eqnarray}
where $ P_{\pm }= P_{x }\pm i P
_{y }. $ The solutions to the Bloch equations (22)
are

\begin{equation}
 \langle P_{z }( t )\rangle = \langle P_{z }(0)\rangle exp(- t /2\tau_{1
}),\eqnum{23}
\end{equation}

\begin{equation}
 \langle P_{\pm }( t )\rangle = \langle P_{\pm }(0)\rangle exp(- t /4\tau_{1 }-
t /4\tau_{0 }),
\eqnum{24}
\end{equation}

Let us now get back to the problem of the E-P-R singlet state
(1) where the two spins are separately interacting with the
external fluctuating magnetic fields. We note that the spin
part of the density matrix corresponding to the initial singlet
state can be written in the following manner:
\begin{eqnarray}
\rho(0)&=&\mid \psi \rangle\langle\psi \mid \nonumber\\
&=& {1 \over 2}\biggl\{
\pmatrix{1 & 0 \cr 0 & 0}_{1}\pmatrix{0 & 0 \cr 0 & 1}_{2} +
\pmatrix{0 & 0 \cr 0 & 1}_{1}\pmatrix{1 & 0 \cr 0 & 0}_{2} \nonumber\\
&&-\pmatrix{0 & 1 \cr 0 & 0}_{1}\pmatrix{0 & 0 \cr 1 & 0}_{2} -
\pmatrix{0 & 0 \cr 1 & 0}_{1}\pmatrix{0 & 1 \cr 0 & 0}_{2}\biggr\},
\nonumber\\ \eqnum{25}
\end{eqnarray}
which can be expressed in terms of the Pauli spin operators
of particles $ 1 $ and $ 2 $ as

\begin{eqnarray}
 \rho (0) &=& { 1\over 4} \biggl( {\bf 1}\otimes{\bf 1} - \sigma_{1}^{x}\sigma
_{2}^{x }- \sigma_{1}^{y}\sigma_{2}^{y
 }- \sigma_{1}^{z}\sigma_{2}^{z }\biggr)\nonumber\\
 &=& { 1\over 4} \biggl( {\bf 1}\otimes{\bf 1} - \sigma_{1 }.\sigma_{2
}\biggr). \eqnum{26}
\end{eqnarray}
This is a convenient form for the initial condition, as we note
that for calculating the correlation between particles $ 1
 $ and $ 2 , $ the spin part of our solution in
Eq. (15) can be written as

\begin{equation}
 \rho_{spin }= { 1\over 4} \biggl({\bf 1}\otimes{\bf 1} + \sigma
_{1 }. P_{1 }( t ) \sigma_{2 }. P_{2 }( t ) \biggr) . \eqnum{27}
\end{equation}

Thus the initial conditions can be regarded as the superposition
of three initial condition corresponding to $ P_{1
 }= x = - P_{2 }; P_{1
}= y = - P_{2 }$ and $ P_{1
 }= z = - P_{2 }. $ Thus, solving
for the time evolution of $ \rho $ involves solving for the
various components of $ P ( $ as in (23), (24)) for each
of the spins with these initial conditions separately and substituting
into Eq.(26). One can easily see that the time dependent
solution is:

\begin{eqnarray}
 \rho ( t ) &=& { 1\over 4} \biggl({\bf 1}\otimes{\bf 1} -(\sigma_{1}^{x}\sigma
_{2}^{x }+ \sigma_{1}^{y}\sigma_{2}^{y })exp(-
 t /4\tau_{0}^{(1) }\nonumber\\
&&- t /4\tau_{1}^{(1) }- t /4\tau_{0}^{(2) }- t /4\tau_{1}^{(2) }) \nonumber\\
&& - \sigma_{1}^{z}\sigma_{2}^{z } \exp
(-  t /2\tau_{0}^{(1) }-t /2\tau_{0}^{(2) }) \biggr), \eqnum{28}
\end{eqnarray}
where $ \tau_{0}^{(1) }, \tau_{0}^{(
2) }, \tau_{1}^{(1) },
\tau_{1}^{(2) }$ are the characteristic
time scales for the particles $ 1 $ and $ 2 . $ We
can now calculate the expectation values required in the Bell
inequalities.

\begin{eqnarray}
 \langle \sigma_{1 }.n_{1 }\sigma_{2 }.n
_{2 }\rangle &=& -(n_{1x }n_{2x }+ n_{1
y }n_{2y })\exp (- { t\over 4} ( 1/\tau_{0}^{(1) }\nonumber\\
&&+1/\tau_{1}^{(1) }+ 1/\tau_{0}^{(2) })) + 1/\tau_{1}^{(2) }\-
n_{1z }n_{2z }\nonumber\\
&&\exp(-t /2\tau_{0}^{(1) }- t /2\tau_{0}^{(2) }).\eqnum{29}
\end{eqnarray}
For simplicity, if we assume that all the time scales are the
same, i.e., $ \tau_{0}^{(1) }=
\tau_{0}^{(2) }= \tau_{1}^{(
1) }= \tau_{1}^{(2) }=\tau_{s }, $ then one can easily see that the
expectation value (29) becomes:

\begin{eqnarray}
 \langle \sigma_{1 }.n_{1 }\sigma_{2 }.n
_{2 }\rangle &=& Trace (\rho ( t )\sigma_{1 }.n_{1 }\sigma_{2 }. n
_{2 })\nonumber\\
 &=& -cos\theta exp(- t /\tau_{s }), \eqnum{30}
\end{eqnarray}
transforming the Bell inequality (3) to

\begin{equation}
 \mid E( a , b )-E( a , c )\mid exp(- t /\tau_{s })\le 1 + E( b ,
c )exp(- t /\tau_{s }). \eqnum{31}
\end{equation}
This shows that as soon as the left-hand side becomes smaller
than unity, the inequalities are satisfied. For the choice of
angles $ \theta ( $ a $ , b ) = \theta ( b , c )
= \pi $ /3 and $ \theta ( $ a $ , c ) = 2\pi $ /3,
the system goes from a violation of the inequality to a nonviolation
at time $ t = -\tau_{s } ln(2/3)$.
Using the solution of the spatial part, i.e., Eq (16), this
means that the quantum correlations are lost when the separation
between the detectors is of the order of $ { 2\epsilon \over m\gamma
} \tau_{s } ln (3/2)$. As mentioned earlier,
the origin of the randomness of the magnetic field is being
attributed to the spatial randomness of the field as seen by
the traversing wave-packets of each particle. It is clear that
for E-P-R correlations to persist for longer and longer times,
one requires isotropy or absence of field fluctuations in larger
and larger portions of space. Indeed, it is important to keep
in mind that this instantaneous quantum propagation of information
from one particle to another is contingent upon the requirement
that the intervening space is free from all field fluctuations
$ - $ a requirement that has to be locally ensured. Another odd
feature worth mentioning is that the decoherence in the spatial
part and the spin part are completely unrelated. Thus, even
when the probability distributions of position of the two particles
have become classical and well-separated, the spin wave-function
may continue to have nonlocal correlations  -  a situation one
would not expect classically.

\section{Summary}

To summarize, in this paper we have studied the effect of the
environment on an E-P-R setup. To simulate an E-P-R setup, we
assume that the two particles of the E-P-R singlet are acted
upon by forces which move them in opposite directions. We find
that environmental influence causes decoherence in the quantum
correlations of the initial singlet state leading to a nonviolation
of Bell's inequalities, thus restoring the classical character
of the correlations. We have also tried to correlate this loss
of coherence to the spatial separation between the two particles
which are additionally coupled to an environment through the
Caldeira-Leggett type of dissipative coupling.


\begin{references}

\bibitem{1} A. Einstein, B. Podolsky, and N. Rosen, Phys. Rev.
47 $ , $ 777 (1935).

\bibitem{2} J. S. Bell, Physics $ 1 , $ 195 (1965).

\bibitem{3} S. J. Freedman and J. F. Clauser, Phys. Rev. Lett.
28 $ , $ 938 (1972).

\bibitem{4} A. Aspect, J. Dalibard and G. Roger, Phys. Rev. Lett.
49 $ , $ 1804 (1982) $ .
 $
\bibitem{5} D. Bohm, Quantum Theory (Prentice-Hall, New York,
1951), reprinted in Quantum Theory and Measurements $ ,
 $ J. A. Wheeler, W. H. Zurek, eds., (Princeton U. P., Princeton,
N. J., 1983).

\bibitem{6} J. F. Clauser and A. Shimony, Rep. Prog. Phys. 41 $ ,
 $ 1881 (1978).

\bibitem{7} F. M. Pipkin in Advances in Atomic and Molecular Physics,
ed. D. R. Bates and B. Bederson (Academic
Press, New York, 1978) Vol. 14, $ p $ 281.

\bibitem{8} C. H. Bennett, G. Brassard, C. $ Cre^{ }$ peau,
R. Josza, A. Peres, and W. K. Wootters,
Phys. Rev. Lett. 70 $ , $ 1895 (1993).

\bibitem{9} A. O. Caldeira and A. J. Leggett, Physica A 121 $ ,
 $ 587 (1983); Phys. Rev. A 31 $ , $ 1057 (1985).

\bibitem{10} A. Venugopalan, D. Kumar and R. Ghosh, Current Science
(1995)

\bibitem{11} R. Kubo, J. Math. Phys. $ 4 , $ 174 (1963).

\bibitem{12} R. Kubo in Lectures in Theoretical Physics, ed.
Brittin and Dunham (Interscience, New York, 1959) Vol. $ 1 ,
p $ 181.
\end{references}
\end{document}